\documentclass[a4paper,11pt]{article}
\usepackage{pos}

\usepackage{multirow}
\usepackage{tabularx}
\usepackage{booktabs}

\newcommand{\OpenLoops}{{\rmfamily\scshape OpenLoops}}

\newcommand{\refeq}[1]{eq.~\eqref{#1}}
\newcommand{\refeqs}[2]{eqs.~\eqref{#1}--\eqref{#2}}
\newcommand{\reffi}[1]{Fig.~\ref{#1}}

\newcommand{\refta}[1]{Table~\ref{#1}}

\newcommand{\refse}[1]{Section~\ref{#1}}

\newcommand{\citere}[1]{Ref.~\cite{#1}}
\newcommand{\citeres}[1]{Refs.~\cite{#1}}

\newcommand{\ie}{i.e.\ }
\newcommand{\eg}{e.g.\ }

\newcommand{\f}[2]{\frac{#1}{#2}}
\newcommand{\sss}[1]{\scriptscriptstyle#1}

\newcommand{\nosss}[1]{#1}

\newcommand{\bea}{\begin{eqnarray}}
\newcommand{\eea}{\end{eqnarray}}
\newcommand{\be}{\begin{equation}}
\newcommand{\ee}{\end{equation}}
\newcommand{\ba}{\begin{align}}
\newcommand{\ea}{\end{align}}
\newcommand{\beas}{\begin{eqnarray*}}
\newcommand{\eeas}{\end{eqnarray*}}
\newcommand{\bes}{\begin{equation*}}
\newcommand{\ees}{\end{equation*}}
\newcommand{\bas}{\begin{align*}}
\newcommand{\eas}{\end{align*}}

\newcommand{\eps}{{\varepsilon}}

\newcommand{\lb}{\left(}
\newcommand{\rb}{\right)}

\newcommand{\Dbar}[1]{{D}_{\nosss{#1}}}
\newcommand{\Db}[2]{{D}_{\nosss{#1}}^{(#2)}}

\newcommand{\momp}[1]{p_{#1}}
\newcommand{\mass}[1]{m_{\nosss{#1}}}

\newcommand{\momq}{\bar{q}}
\newcommand{\tilq}{\tilde{q}}

\newcommand{\calC}{\mathcal{C}}

\newcommand{\calN}{\mathcal{N}}
\newcommand{\barN}{\bar{\mathcal{N}}}

\newcommand{\barM}{\bar{\mathcal{M}}}
\newcommand{\topo}{\mathcal{\tau}}
\newcommand{\topoconf}{\vec{n}}
\newcommand{\topoconfdelta}{\Delta\vec{n}}
\newcommand{\nind}[1]{n_{\sss{#1}}}
\newcommand{\tilnind}[1]{m_{\sss{#1}}}

\newcommand{\calV}{\mathcal{V}}

\newcommand{\fullampbar}[2]{{\barM}_{{#1},{#2}}}
\newcommand{\colfac}[2]{{C}_{{#1},{#2}}}
\newcommand{\vertex}[1]{\calV_{#1}}

\newcommand{\rd}{\mathrm d}

\newcommand{\nextmom}{N_p}

\definecolor{bluemar}{rgb}{0,0,.5}
\definecolor{redmar}{rgb}{.8,0,0}
\definecolor{greenmar}{rgb}{0,.5,0}

\graphicspath{{./pdfdiagrams/},{./figs/}}

\title{Recursive reduction of two-loop tensor integrals}

\author*[a,b]{Fabian Lange}
\author[a,b]{Max F.~Zoller}

\affiliation[a]{Physik-Institut, Universität Zürich,\\
Winterthurerstrasse 190, 8057 Zürich, Switzerland}

\affiliation[b]{PSI Center for Neutron and Muon Sciences,\\
5232 Villigen PSI, Switzerland}

\emailAdd{fabian.lange@physik.uzh.ch}
\emailAdd{max.zoller@physik.uzh.ch}

\abstract{In order to meet the precision requirements for the LHC and future colliders, next-to-next-to-leading order corrections to a wide range of processes are essential, making general automated tools highly desirable.
Extending the strategy of the widespread one-loop program \OpenLoops{} to two loops, there are three major ingredients: process-dependent tensor coefficients, tensor integrals, and process-independent counterterms.
In these proceedings, we focus on the second part and present a new recursive algorithm to reduce arbitrary two-loop tensor integrals to scalar integrals numerically.}

\FullConference{17th International Symposium on Radiative Corrections: Applications of Quantum Field Theory to Phenomenology (RADCOR2025)\\
5-10 October 2025\\
Puri, India\\}

\begin{document}
\maketitle

\section{Introduction}

Perturbative scattering amplitudes are a key ingredient for Monte Carlo simulations of collider processes.
Tree and one-loop amplitudes have been available from fully automated numerical tools, such as \OpenLoops{}~\cite{Cascioli:2011va,Buccioni:2019sur}, for many years.
In order to meet the precision requirements of the LHC and future colliders, an extension to two loops is essential.
The key idea in the \OpenLoops{} framework is to split the calculation into three major ingredients: process-dependent tensor coefficients, tensor integrals, and process-independent counterterms.
We give a brief overview over this approach in \refse{se:openloops}, before focusing on the tensor integral reduction.
This topic got some attention in the literature in recent years, see \eg \citeres{Chen:2019wyb,Peraro:2019cjj,Peraro:2020sfm,Anastasiou:2023koq,Goode:2024mci,Goode:2024cfy,Bevilacqua:2025xms}.
We have developed a new recursive algorithm that integrates particularly well into the numerical \OpenLoops{} framework.
We introduce this algorithm first for the one-loop case in \refse{se:red1l}, before extending it to two loops in \refse{se:red2l} and showing some preliminary validation and performance tests in \refse{se:validation}.

\section{Automated calculations with tensor integral decomposition}
\label{se:openloops}

We compute amplitudes in the 't~Hooft--Veltman scheme~\cite{tHooft:1972tcz}, where external wave functions and momenta are four-dimensional, while loop momenta $\momq_i$, metric tensors $\bar{g}^{\bar\mu\bar\nu}$ and Dirac matrices $\bar\gamma^{\bar\mu}$ inside loops are defined in $D=4-2\eps $ dimensions in order to regularise divergences in loop integrals.
We denote these $D$-dimensional quantities with a bar, their projection to four dimensions without a bar, and the $(D-4)$-dimensional difference between those with a tilde, \ie
\bea
\momq_i = q_i + \tilde{q}_i , \quad
\bar{g}^{\bar\mu\bar\nu} = {g}^{\mu\nu} + \tilde{g}^{\tilde\mu\tilde\nu} , \quad \text{and} \quad
\bar{\gamma}^{\bar\mu} = {\gamma}^{\mu} + \tilde{\gamma}^{\tilde\mu} .
\eea

\subsection{One loop}

\begin{figure}[bp]
  \centering
  \includegraphics[width=0.2\textwidth]{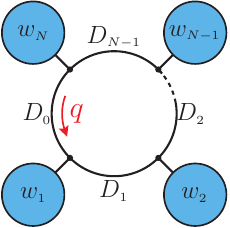}
  \caption{
    \label{fig:oneloopdia}
    General structure of a one-loop diagram.
    The blue blobs denote external subtrees.
  }
\end{figure}

An individual one-loop diagram $\Gamma$ as depicted in \reffi{fig:oneloopdia} has the form
\be
\fullampbar{1}{\Gamma} =
\colfac{1}{\Gamma}
\int\!\rd\momq\,
\f{\barN(\momq)}
{\mathcal{D}(\momq)}\,,
\label{eq:intro_amp_one}
\ee
where $\colfac{1}{\Gamma}$ is a colour factor.
The integration measure in loop-momentum space is abbreviated as
\bea
\int\!\rd\momq & = & \mu^{2\eps}\int\!\f{\rd^{^D}\! \bar
q}{(2\pi)^{^D}}\,,
\eea
where $\mu$ is the scale of dimensional regularisation.
The denominator is a product
\be
\mathcal{D}(\momq)=\prod\limits_{a=0}^{N-1} \Dbar{a}(\momq)
\ee
of $N$ propagator denominators
\be
\Dbar{a}(\momq) = (\momq+p_{a})^2-m_{a}^2 {}
\ee
with masses $m_{a}$ and external momenta $p_{a}$ flowing through the propagator.
For the computation within the numerical \OpenLoops{} framework we decompose the numerator
into a four-dimensional part $\calN(q)=\barN(\momq)|_{D\to 4}$
and a remainder $\tilde{\calN}$, which is of order $(D-4)$ and contributes only through its interplay with integral divergences. We further decompose the numerator into loop-momentum tensors
\be
\calN(q) = \sum\limits_{r=0}^R
\calN_{\mu_1\ldots\mu_r}
q^{\mu_1}\ldots q^{\mu_r}
\label{eq:intro_tdec_one}
\ee
with maximum tensor rank $R$.
This leads to a decomposition of the amplitude
\be
\fullampbar{1}{\Gamma} =
\colfac{1}{\Gamma} \lb \sum\limits_{r=0}^R \calN_{\mu_1\ldots\mu_r}
I_{\topo}^{\mu_1\ldots\mu_{r}}
+
\int\!\rd\momq\,
\f{\tilde{\calN}(\momq)}
{\mathcal{D}(\momq)}\rb \label{eq:1lamp_dec}
\ee
with one-loop rank-$r$ tensor integrals of the general form
\bea
I^{\mu_1\cdots\mu_r}_{\topo}(\topoconf) &=& \int\!\rd\momq\,
\f{q^{\mu_1}\cdots q^{\mu_{r}} \,\lb\tilq^2\rb^\tilnind{}}{
\prod\limits_{b=0}^{N-1} \lb \Dbar{b}(\momq) \rb^{\nind{b}}}.
\label{eq:TIintegranl1l_def}
\eea
These are characterised by the set of propagator denominators $\{\Dbar{a}\}$ of $\Gamma$, which we denote as the topology $\tau$, and a given set of integer indices
\be
\topoconf=(\tilnind{},\nind{0},\ldots,\nind{N-1})
\label{eq:n_config_def}
\ee
as a configuration of that topology.
The index $\tilnind{}$ marks the power of the $(D-4)$-dimensional contribution $\Dbar{-2}=\tilq^2$ in the numerator of the integral, which for the integral in \refeq{eq:1lamp_dec} is $\tilnind{}=0$, while the other indices are typically $\nind{a}\in\{0,1,2\}$.

The tensor coefficients $\calN_{\mu_1\ldots\mu_r}$ are constructed through a numerical recursion~\cite{vanHameren:2009vq,Cascioli:2011va,Buccioni:2017yxi} exploiting the factorisation of amplitudes into process-independent building blocks depending only on the Feynman rules of a given model.
The contribution of the $(D-4)$-dimensional numerator parts $\tilde{\calN}$ are reconstructed through the insertion of process-independent rational counterterms~\cite{Ossola:2008xq,Draggiotis:2009yb,Garzelli:2009is,Pittau:2011qp} into tree-level diagrams together with the usual ultraviolet (UV) counterterms in the chosen renormalisation scheme.
In the public \OpenLoops{}~2 program~\cite{Buccioni:2019sur}, the tensor integrals are reduced to scalar master integrals with an on-the-fly reduction algorithm~\cite{Buccioni:2017yxi} and external tools~\cite{Denner:2016kdg,Ossola:2007ax}.

\subsection{Two loops}

\begin{figure}[htbp]
  \centering
  \includegraphics[width=0.45\textwidth]{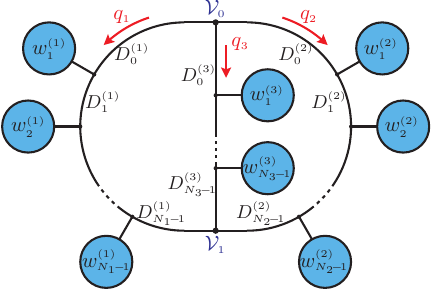}
  \caption{
    \label{fig:twoloopdia}
    General structure of an irreducible two-loop diagram.
    The blue blobs denote external subtrees.
    In general $\vertex{0}, \vertex{1}$ can be quartic vertices, in which case an external subtree is also attached there.
  }
\end{figure}

The amplitude of a generic two-loop diagram $\Gamma$ has the form
\be
\fullampbar{2}{\Gamma} =
\colfac{2}{\Gamma}
\int\!\rd\momq_1\int\!\rd\momq_2\,
\f{\barN(\momq_1,\momq_2)}
{\mathcal{D}(\momq_1,\momq_2)} \label{eq:intro_amp_two}\,,
\ee
where $\colfac{2}{\Gamma}$ is a colour factor, and the denominator
\be
\mathcal{D}(\momq_1,\momq_2)= \prod\limits_{i=1}^3 \prod\limits_{a=0}^{N_i-1} \Db{a}{i}(\momq_i) \quad \text{with} \quad \Db{a}{i}(\momq_i) = (\momq_i+p_{ia})^2-m_{ia}^2 ,
\label{eq:den_prod_twoloop}
\ee
which depend on one of the independent loop momenta $\momq_1,\momq_2$ or the linear combination $\momq_3=-(\momq_1+\momq_2)$ as well as a masses $m_{ia}$ and an external momenta $p_{ia}$.
We choose the loop momenta in such a way that the number of propagators in \refeq{eq:den_prod_twoloop} is ordered as $N_1 \geq N_2 \geq N_3$.
For $N_3=0$ the two-loop diagram is reducible, \ie it factorises into two one-loop integrals.
An efficient algorithm for the construction of such amplitudes was discussed in \citere{Pozzorini:2022ohr}.
The more challenging case of an irreducible diagram with $N_3\geq 1$ is depicted in \reffi{fig:twoloopdia}.
Its integrand factorises into three propagator chains $\calC_i$, each depending on a single loop momentum, connected by two vertices $\vertex{0},\vertex{1}$, which in general depend on both independent loop momenta $q_1,q_2$.

In our approach, we again decompose the numerator of such a diagram into a four-dimensional part $\calN(q_1,q_2)=\barN(\momq_1,\momq_2)|_{D\to 4}$ and a remainder $\tilde{N}(\momq_1,\momq_2)$, which is of order $(D-4)$ and contributes only through its interplay with integral divergences.
Similarly as in~\refeq{eq:intro_tdec_one}, we apply a tensor decomposition to the four-dimensional numerator
\be
\calN(q_1,q_2) =
\sum\limits_{r=0}^{R_1}\sum\limits_{s=0}^{R_2(r)}\calN_{\mu_1\ldots\mu_r,\,\nu_1\ldots\nu_s}\,
q_1^{\mu_1}\ldots q_1^{\mu_r} q_2^{\nu_1}\ldots
q_2^{\nu_s}\, \label{eq:intro_tdec_two}
\ee
with upper bounds $R_i$ for the tensor ranks in the loop momenta $q_i$.
This decomposition leads to
\be
\fullampbar{2}{\Gamma} =
\colfac{2}{\Gamma} \lb \sum\limits_{r=0}^{R_1}\sum\limits_{s=0}^{R_2(r)}\calN_{\mu_1\ldots\mu_r,\,\nu_1\ldots\nu_s}\,
I_{\topo}^{\mu_1\ldots\mu_{r};\nu_1\ldots\nu_s}
+
\int\!\rd\momq_1\int\!\rd\momq_2\,
\f{\tilde{\calN}(\momq_1,\momq_2)}
{\mathcal{D}(\momq_1,\momq_2)}\rb
\label{eq:ampdia2L}
\ee
with two-loop tensor integrals of the form
\bea I_\tau^{\mu_1\ldots\mu_{r};\nu_1\ldots\nu_{s}}(\topoconf)
&=& \int\!\rd\momq_1\!\int\!\rd\momq_2
\f{q_1^{\mu_1}\cdots q_1^{\mu_{r}}\,q_2^{\nu_1}\cdots q_2^{\nu_{s}} \prod\limits_{i=1}^3 (\tilq_i^2)^{m_i}}{
 \prod\limits_{i=1}^3 \prod\limits_{a=0}^{N_i-1} \left[\Db{a}{i}(\momq_i)\right]^{\nind{ia}}} \quad\Big|_{q_3\to -(q_1+q_2)}{},
\label{eq:TIdia2l-ind}
\eea
where $\tau$ again labels a given topology, and the set of indices is defined as
\be
\topoconf = (\vec{m},\topoconf_1,\topoconf_2,\topoconf_3) \label{eq:n_config_def_2l}
\ee
with $\vec{m} = (\tilnind{1},\tilnind{2},\tilnind{3})$ and $\topoconf_i = (\nind{i0},\ldots,\nind{i\,N_i-1})$.
As in the one-loop case $\vec{m}=0$ and usually $\nind{ia}\in\{0,1,2\}$ for the integral in \refeq{eq:ampdia2L}.

A highly efficient and numerically stable algorithm to recursively construct the tensor coefficients $\calN_{\mu_1\ldots\mu_r,\,\nu_1\ldots\nu_s}$ was presented in \citere{Pozzorini:2022ohr}.
Similarly as in the one-loop case, the contributions associated with the interplay of $\tilde{\calN}$ with poles
can be reconstructed by means of two-loop rational counterterms~\cite{Pozzorini:2020hkx,Lang:2020nnl,Lang:2021hnw} together with the usual UV counterterms.
This approach is fully established for the interplay of $\tilde{\calN}$ with UV poles, while the rational terms of infrared (IR) origin are still under investigation~\cite{Zhang:2022rft}.
In these proceedings, we focus on the tensor integrals \eqref{eq:TIdia2l-ind}.

\section{Recursive reduction of one-loop tensor integrals}\label{se:red1l}

\subsection{Integrand-level reduction by one rank}\label{se:red1lrank2}

A four-dimensional loop-momentum tensor of rank $2$ can be reduced with the exact identity~\cite{delAguila:2004nf}
\bea q^\mu q^\nu
 &=&
 \sum\limits_{a=-2}^{3} \lb
  A^{\mu\nu}_{a} +B^{\mu\nu}_{a,\lambda} \,q^{\lambda} \rb \Dbar{a}(q) ,
\label{eq:qmuqnuredfinal}
\eea
where the coefficients $A^{\mu\nu}_{a}$ and $B_{a,\lambda}^{\mu\nu}$ on the RHS are composed of three independent external momenta $p_{1},p_{2},p_{3}$, metric tensors, and the masses $\mass{a}$ of the propagators
\bea
 \Dbar{a}(\momq) = \begin{cases}
                        (\momq + \momp{a})^2-\mass{a}^2 & \text{ for } a \geq 0 , \\
                        1 & \text{ for } a = -1 ,\\
                        \tilq^2 & \text{ for } a = -2 .
                      \end{cases}
                      \label{eq:generalprop}
\eea
With this we can reduce the tensor integrand by one rank and obtain
\bea
\f{q^{\mu_1}\cdots q^{\mu_{r}}}{
\prod\limits_{b=0}^{N-1} \Dbar{b}(\momq) }
&=&
\sum\limits_{a=-2}^{3}
\f{\lb A^{\mu\nu}_{a} + B^{\mu\nu}_{a,\lambda}q^{\lambda}\rb
q^{\mu_3}\cdots q^{\mu_{r}}}{
\prod\limits_{b=-2}^{N-1} \lb \Dbar{b}(\momq)\rb^{1-\delta_{ab}} }
.
\label{eq:TIintegrandred1l}
\eea
The RHS of \refeq{eq:TIintegrandred1l} consists of loop momentum tensors of rank $r-1$ and $r-2$ and configurations with the original set of propagators and sets where one propagator denominator has been canceled.
As long as at least four $\Dbar{a}$ remain, \refeq{eq:qmuqnuredfinal} can be used recursively.
In \OpenLoops{}~2~\cite{Buccioni:2019sur} an on-the-fly version of this method~\cite{Buccioni:2017yxi} was implemented, where construction steps for the tensor coefficients of a one-loop amplitude are interleaved with reduction steps of the form of \refeq{eq:TIintegrandred1l} in order to keep the tensor rank $\leq 2$ throughout the calculation.

However, the integrand-level identity~(\ref{eq:TIintegrandred1l}) can be also be applied to tensor integrals with arbitrary exponents of the propagator denominators.
Applying the reduction identity~(\ref{eq:qmuqnuredfinal}) as in \refeq{eq:TIintegrandred1l} yields
\bea
I^{\mu_1\cdots\mu_r}_{\topo}(\topoconf)
&=&
\sum\limits_{a=-2}^{3} \left[
A^{\mu_1\mu_2}_{a}
I^{\mu_3\cdots\mu_r}_{\topo}(\topoconf+\topoconfdelta_a)
+
B^{\mu_1\mu_2}_{a,\lambda}
I^{\lambda\mu_3\cdots\mu_r}_{\topo}(\topoconf+\topoconfdelta_a)
\right]
\label{eq:TIintegranl1l_red_ind}
\eea
with integer indices
\be
\topoconfdelta_{a}=(\delta_{-2,a},-\delta_{0,a},\ldots,-\delta_{N-1,a}).
\ee
The rank-$(r-1)$ and rank-$(r-2)$ integrals on the RHS can again be reduced with \refeq{eq:qmuqnuredfinal}. Allowing for negative indices in topology configurations, we can use the same set of four propagators $\Dbar{0}\ldots\Dbar{3}$ in all steps of this recursion.
Only the indices $\tilnind{},\nind{0},\nind{1},\nind{2},\nind{3}$ are changed throughout the reduction.

\subsection{Reduction of one-loop tensor integrals to rank one}\label{se:red1lrankr}

A rank-$r$ tensor in a single loop momentum $q$ can be reduced to a rank-$1$ tensor by recursive application of
\refeq{eq:qmuqnuredfinal}. The result is expressed in terms of reconstructed denominator products
\be
    \Dbar{\Omega} =
    \prod\limits_{a=-2}^{3}\lb\Dbar{a}(\momq)\rb^{\omega_{a}},
    \label{eq:D_Gammar_def_1l}
\ee
where $\Omega = \{\omega_{-2}, \dots, \omega_{3}\}$ is a set of integer exponents
with $\sum\limits_a \omega_a = r-1$. The final result of such a recursive reduction has the form
\begin{equation}
  q^{\mu_1} \cdots q^{\mu_r} = \sum\limits_{\Omega\in\Gamma_r} \left[ A^{\mu_1 \cdots \mu_r}_{\Omega} + B^{\mu_1 \cdots \mu_r}_{\Omega, \lambda} q^\lambda \right] \Dbar{\Omega} ,
  \label{eq:q-reduction-rank-r}
\end{equation}
where the sum is performed over the set of all possible denominator products described by a set of six non-negative integer indices. We define
\bea
  \Gamma_r &=& \Big\{  \Omega = \{\omega_{-2}, \dots, \omega_{3}\} \;\;\Big|\;\;
  \omega_a \in \mathbb{N}_0,\;\;
  \sum\limits_a \omega_a = r-1
  \Big\}. \label{eq:Gammar_def}
\eea
Applying \refeq{eq:q-reduction-rank-r} to the tensor integral~(\ref{eq:TIintegranl1l_def})
gives
\bea
I^{\mu_1\cdots\mu_r}_{\topo}(\topoconf)
&=&
\sum\limits_{\Omega\in\Gamma_r}
\left[ A^{\mu_1 \cdots \mu_r}_{\Omega} I_{\topo}(\topoconf+\topoconfdelta_\Omega)
+ B^{\mu_1 \cdots \mu_r}_{\Omega, \lambda} I^{\lambda}_{\topo}(\topoconf+\topoconfdelta_\Omega)
\right],
\label{eq:TIintegranl1l_fullred_ind}
\eea
where we define the configuration shift vector with the same dimension as $\topoconf$ as
\be
\topoconfdelta_\Omega = (\omega_{-2},-\omega_{0},-\omega_{1},-\omega_{2},-\omega_{3},
0,\ldots,0).
\label{eq:n_configshift}
\ee
Since only $\tilq^2$ and the first four propagator denominators partake in the recursive reduction, all other denominators are not shifted.
For a general tensor rank $r\geq 3$, the coefficients in \refeq{eq:q-reduction-rank-r} can be computed recursively from coefficients of rank $r-1$ and of rank $2$ as
\begin{equation}
  \begin{split}
    A^{\mu_1 \cdots \mu_r}_{\Omega} &= \sum\limits_{a =-2}^3 \theta(\Omega-\Omega_a)\,
    B^{\mu_1 \cdots \mu_{r-1}}_{\Omega-\Omega_a, \lambda^\prime} A^{\lambda^\prime \mu_r}_{\Omega_a}, \\
    B^{\mu_1 \cdots \mu_r}_{\Omega, \lambda} &= \sum\limits_{a =-2}^3 \theta(\Omega-\Omega_a)\left[ B^{\mu_1 \cdots \mu_{r-1}}_{\Omega-\Omega_a, \lambda^\prime} B^{\lambda^\prime \mu_r}_{\Omega_a, \lambda} + A^{\mu_1 \cdots \mu_{r-1}}_{\Omega-\Omega_a} \delta^{\mu_r}_\lambda \delta_{a,-1} \right]
  \end{split}
  \label{eq:A-B-recursion}
\end{equation}
with
\bea
  \Omega - \Omega_a &=& (\omega_{-2},\ldots,\omega_{a}-1,\ldots\omega_3).
\eea
The theta function restricts the sum to values of $a$ for which $\omega_a \geq 1$, \ie
\be
  \theta(\Omega-\Omega_a) = \theta(\omega_a-1) \quad\quad\text{ for }\quad \Omega=\{\omega_{-2},\ldots,\omega_3\}
\ee
with the usual Heaviside step function $\theta$ on the RHS.
We will proof these formulae in a forthcoming publication~\cite{Lange:2026XXX}.
This recursive construction is well-suited to our purpose, since for a Feynman diagram with maximum loop momentum tensor rank $R$ in the numerator, tensor integrals with all ranks $r=0,\ldots,R$ appear, and hence all the reduction coefficients up to rank $R$ need to be constructed. In particular, we use the same rank-$2$ coefficients in every step.
The main advantage with this type of recursive reduction is that we avoid large systems of equations altogether.

\subsection{Reduction to rank zero}\label{se:red1ltorank0}

Applying the reduction~(\ref{eq:TIintegranl1l_fullred_ind}) with the reduction coefficients~(\ref{eq:A-B-recursion}) to a tensor integral~(\ref{eq:TIintegranl1l_def}) of rank $r$ leaves us with integrals of rank 1 and rank 0 in $q$.
In addition to the scalar propagator denominators $\Dbar{0},\ldots,\Dbar{N-1}$ they can also feature powers of $\Dbar{-2}=\tilde{q}^2$,
\bea
I^{(\lambda)}_{\topo}(\topoconf)
&=& \int\!\rd\momq\,
\f{\lb\tilde{q}^2\rb^{m}}{
\prod\limits_{b=0}^{N-1} \lb \Dbar{b}(\momq) \rb^{\nind{b}}}
\cdot
\begin{cases}
  1 , \\
  q^\lambda ,
\end{cases}
\label{eq:1l_qtildeint}
\eea
with the configuration $\topoconf$ as defined in \refeq{eq:n_config_def} and the index $m=0,\ldots,\lfloor\f{r}{2}\rfloor$.

$\tilde{q}^2$ is a scalar in the $(D-4)$-dimensional subspace $V_{D-4}$ of the full $D$-dimensional momentum space $V_D$ (but not in $V_D$), and $q^\lambda$ is a vector in the four-dimensional subspace $V_{4}$. The two subspaces are disjoint and $V_D = V_4 + V_{D-4}$.
Hence, their reduction can be factorised even using integral-level identities, such as Passarino-Veltman reduction~\cite{Passarino:1978jh}.
With this, we arrive at the reduction formula
\bea
I^{\mu_1\cdots\mu_r}_{\topo}(\topoconf)
&=&
\sum\limits_{\Omega\in\Gamma_r}
\sum\limits_{k=-1}^{\nextmom}
C^{\mu_1 \cdots \mu_r}_{\Omega,k} I_{\topo}(\topoconf+\topoconfdelta_\Omega+\topoconfdelta_k)
\label{eq:TIintegranl1l_fullred_toRank0}
 \eea
with
\be
C^{\mu_1 \cdots \mu_r}_{\Omega,k} = B^{\mu_1 \cdots \mu_r}_{\Omega, \lambda} C_{k}^\lambda
\;+\; \delta_{-1,k} \,A^{\mu_1 \cdots \mu_r}_{\Omega} ,
\label{eq:TIintegranl1l_fullred_toRank0_coeff_ck}
\ee
where the coefficients $C_{k}^\lambda$ are computed with the Passarino-Veltman reduction.
Each of the $C_{k}^\lambda$ comes together with the configuration shift
\bea
  \topoconfdelta_k &=& (0,-\delta_{k,0},\ldots,-\delta_{k,N-1}) .
\eea

The remaining integrals are of the form of the first line of \refeq{eq:1l_qtildeint} and fall into two groups:
Integrals with $m=0$ are scalar integrals that can be computed with standard tools.
The second group are integrals with $m>0$.
Since $\tilde{q}^2$ is of order $(D-4)$, these integrals can only contribute to the finite part of the result through their interplay with $\f{1}{(D-4)^n}$ poles of the loop integral.
At one-loop level these contributions are known as rational terms of type $R_1$~\cite{Ossola:2008xq,Draggiotis:2009yb,Garzelli:2009is,Pittau:2011qp}.

\section{Recursive reduction of two-loop tensor integrals}\label{se:red2l}

For the case of more than one independent loop momenta $q_i$, we can apply~\refeq{eq:qmuqnuredfinal} to each loop momentum chain with four independent propagator denominators.
To this end, and in order to represent the irreducible scalar products, it can be necessary to extend and/or reexpress the set of propagator denominators in a loop integral stemming from a single Feynman diagram or a sum of diagrams.
Applying \refeq{eq:qmuqnuredfinal} to a tensor of rank $r$ in the loop momentum $q_1$ and rank $s$ in the loop momentum $q_2$ gives
\bea
q_1^{\mu_1} \cdots q_1^{\mu_r}q_2^{\nu_1} \cdots q_2^{\nu_s} = \sum\limits_{\Omega\in\Gamma_r} \sum\limits_{\Omega'\in\Gamma_s}
\left[ A^{\mu_1 \cdots \mu_r}_{1,\Omega} +
B^{\mu_1 \cdots \mu_r}_{1,\Omega, \lambda} q_1^{\lambda} \right]
\left[ A^{\nu_1 \cdots \nu_s}_{2,\Omega'} +
B^{\nu_1 \cdots \nu_s}_{2,\Omega', \lambda'} q_2^{\lambda'} \right]
\Dbar{\Omega}^{(1)}\Dbar{\Omega'}^{(2)},
  \label{eq:q-reduction-rank-r-2l}
\eea
where $\Dbar{\Omega}^{(i)}$ are defined as in \refeq{eq:D_Gammar_def_1l} with the propagators of \refeq{eq:generalprop} now carrying an index $(i)$ to denote the corresponding loop momentum, cf.\ \refeq{eq:den_prod_twoloop}.
With this reduction any two-loop integrand can be reduced to a sum of rank-$0$, rank-$1$ and mixed (rank-$1$) $\times$ (rank-$1$) integrands in the four-dimensional indices, using recursively derived reduction coefficients of one-loop complexity. At the level of tensor integrals \eqref{eq:TIdia2l-ind} this reduction reads\footnote{For clarity we use the semicolon separating the tensor indices of $q_1$ and $q_2$ in all integrals with a rank larger than $0$, but leave it out in integrals with total rank $0$.}
%
%
\bea I_\tau^{\mu_1\ldots\mu_{r};\nu_1\ldots\nu_{s}}(\topoconf)
&=&
\sum\limits_{\Omega\in\Gamma_r} \sum\limits_{\Omega'\in\Gamma_s} \Big[
A^{\mu_1 \cdots \mu_r}_{1,\Omega} A^{\nu_1 \cdots \nu_s}_{2,\Omega'}
I_{\topo}(\topoconf+\topoconfdelta_{\Omega,\Omega'}) \nonumber  +
A^{\nu_1 \cdots \nu_s}_{2,\Omega'} B^{\mu_1 \cdots \mu_r}_{1,\Omega, \lambda}
I^{\lambda;}_{\topo}(\topoconf+\topoconfdelta_{\Omega,\Omega'}) \nonumber \\ && +
A^{\mu_1 \cdots \mu_r}_{1,\Omega} B^{\nu_1 \cdots \nu_s}_{2,\Omega', \lambda'}
I^{;\lambda'}_{\topo}(\topoconf+\topoconfdelta_{\Omega,\Omega'}) +
B^{\mu_1 \cdots \mu_r}_{1,\Omega, \lambda} B^{\nu_1 \cdots \nu_s}_{2,\Omega', \lambda'}
I^{\lambda;\lambda'}_{\topo}(\topoconf+\topoconfdelta_{\Omega,\Omega'}) \Big] ,
\label{eq:TIdia2l-red_rank1}
\eea
where we define the configuration shift vector with the same dimension as $\topoconf$ in terms of the components $\omega_{a}$ of $\Omega$ and the components $\omega'_{a}$ of $\Omega'$ as $\topoconfdelta_{\Omega,\Omega'} = (\Delta\vec{m},\topoconfdelta_1,\topoconfdelta_2,\vec{0})$ with
\bea
\Delta\vec{m} \ \ = \ \ (\omega_{-2},\omega'_{-2},0), \quad \topoconfdelta_1 &=& (-\omega_{0},-\omega_{1},-\omega_{2},-\omega_{3}, 0,\ldots,0),\nonumber\\
\topoconfdelta_2 &=& (-\omega'_{0},-\omega'_{1},-\omega'_{2},-\omega'_{3}, 0,\ldots,0).
\label{eq:n_configdelta_def_2l_ind}
\eea
Note that both $\topoconf_3$ and $m_3$ stay unchanged by this part of the reduction.

Similarly as in the one-loop case in \refeq{eq:1l_qtildeint}, we are left with integrals
\bea
  I^{(\lambda);(\lambda')}_{\topo}(\topoconf) &=&
  \int\!\rd\momq_1\!\int\!\rd\momq_2
  \f{\lb\tilde{q}_1^2\rb^{m_1}\lb\tilde{q}_2^2\rb^{m_1}}{
  \prod\limits_{i=1}^3 \prod\limits_{a=0}^{N_i-1} \Db{a}{i}(\momq_i) }
  \cdot
  \begin{cases}
    1 , \\
    q_1^{\lambda} , \\
    q_2^{\lambda'} , \\
    q_1^{\lambda} q_2^{\lambda'} \\
  \end{cases}
\eea
of at most rank $1$ along each chain.
Again using Passarino-Veltman reduction~\cite{Passarino:1978jh}, we can construct reduction coefficients $C_{k}^{(1)\,\lambda}$, $C_{k}^{(2)\,\lambda'}$, and $C_{k}^{(3)\,\lambda\lambda'}$ and arrive at the reduction formula
\bea
I_\tau^{\mu_1\ldots\mu_{r};\nu_1\ldots\nu_{s}}(\topoconf)
&=&
\sum\limits_{\Omega\in\Gamma_r} \sum\limits_{\Omega'\in\Gamma_s}
\Bigg[
  C^{(0)\,\mu_1\ldots\mu_{r};\nu_1\ldots\nu_{s}}_{\Omega,\Omega'} I_{\topo}(\topoconf+\topoconfdelta_{\Omega,\Omega'}) \nonumber\\
  &&\hspace*{5em} + \sum\limits_{i=1}^{2}\sum\limits_{k=-1}^{\nextmom}
  C^{(i)\,\mu_1\ldots\mu_{r};\nu_1\ldots\nu_{s}}_{\Omega,\Omega',k} I_{\topo}(\topoconf+\topoconfdelta_{\Omega,\Omega'}+\topoconfdelta^{(i)}_{k})\nonumber\\
  &&\hspace*{5em}
  + \sum\limits_{k=-2}^{3+4\nextmom+\nextmom^2}
  C^{(3)\,\mu_1\ldots\mu_{r};\nu_1\ldots\nu_{s}}_{\Omega,\Omega',k} I_{\topo}(\topoconf+\topoconfdelta_{\Omega,\Omega'}+\topoconfdelta^{(3)}_{k})
\Bigg]
\label{eq:TIintegranl2l_fullred_toRank0}
\eea
with
\bea
  C^{(0)\,\mu_1\ldots\mu_{r};\nu_1\ldots\nu_{s}}_{\Omega,\Omega'} &=& A^{\mu_1 \cdots \mu_r}_{1,\Omega} A^{\nu_1 \cdots \nu_s}_{2,\Omega'} ,\nonumber\\
  C^{(1)\,\mu_1\ldots\mu_{r};\nu_1\ldots\nu_{s}}_{\Omega,\Omega',k} &=& A^{\nu_1 \cdots \nu_s}_{2,\Omega'} B^{\mu_1 \cdots \mu_r}_{1,\Omega, \lambda} C_{k}^{(1)\,\lambda} ,\nonumber\\
  C^{(2)\,\mu_1\ldots\mu_{r};\nu_1\ldots\nu_{s}}_{\Omega,\Omega',k} &=& A^{\mu_1 \cdots \mu_r}_{1,\Omega} B^{\nu_1 \cdots \nu_s}_{2,\Omega', \lambda'} C_{k}^{(2)\,\lambda'} ,\nonumber\\
  C^{(3)\,\mu_1\ldots\mu_{r};\nu_1\ldots\nu_{s}}_{\Omega,\Omega',k} &=& B^{\mu_1 \cdots \mu_r}_{1,\Omega, \lambda} B^{\nu_1 \cdots \nu_s}_{2,\Omega', \lambda'} C_{k}^{(3)\,\lambda\lambda'}
\label{eq:TIintegranl2l_fullred_toRank0_coeff_ck}
\eea
and configuration shift vectors $\topoconfdelta^{(1)}_{k}$, $\topoconfdelta^{(2)}_{k}$ changing only indices of the first and second loop momentum chain, respectively, and $\topoconfdelta^{(3)}_{k}$ acting on all three chains.
We refer to our forthcoming publication for more details~\cite{Lange:2026XXX}.

With \refeqs{eq:TIintegranl2l_fullred_toRank0}{eq:TIintegranl2l_fullred_toRank0_coeff_ck} we achieved a complete reduction of the two-loop tensor integrals.
In a next step, they have to be contracted with the tensor coefficients $\calN_{\mu_1\ldots\mu_r,\,\nu_1\ldots\nu_s}$ to obtain the amplitude in \refeq{eq:ampdia2L}.
While \refeq{eq:TIintegranl2l_fullred_toRank0} allows us to construct every single component of the tensor integrals, which can checked with other methods, this involves a huge number of tensor components and contractions.
However, to obtain the amplitude, it is not necessary to compute the tensor components of the tensor integrals individually.
We can first contract the tensor coefficients $\calN_{\mu_1\ldots\mu_r,\,\nu_1\ldots\nu_s}$ with the reduction coefficients along the first chain, \ie
\be
  \calN_{\mu_1\ldots\mu_r,\,\nu_1\ldots\nu_s} A^{\mu_1 \cdots \mu_r}_{1,\Omega} \quad \text{and} \quad \calN_{\mu_1\ldots\mu_r,\,\nu_1\ldots\nu_s} B^{\mu_1 \cdots \mu_r}_{1,\Omega, \lambda} ,
  \label{eq:amp-mode}
\ee
instead of constructing \refeq{eq:TIintegranl2l_fullred_toRank0_coeff_ck} and perform the remaining steps on this contracted object.
This has the advantage that a large number of components is contracted in the first step and does not appear anymore at later stages.
Of course, there are many ways to arrange the order of contractions.
We will discuss this in more detail in a forthcoming publication~\cite{Lange:2026XXX}.
As a first glimpse, we compare the \emph{tensor integral mode} based on \refeq{eq:TIintegranl2l_fullred_toRank0_coeff_ck} with a first version of the \emph{amplitude mode} following the idea of \refeq{eq:amp-mode} in the next section.

\section{Validation and performance tests of the two-loop reduction}
\label{se:validation}

\begin{figure}[htbp]
  \centering
  \includegraphics[width=0.2\textwidth]{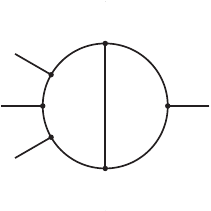}
  \caption{
    Massless two-loop pentagon-triangle topology used to test our implementation.
  }
  \label{fig:pentagon-triangle}
\end{figure}

As a first check of our \texttt{Fortran} implementation we consider the massless $2\to2$ pentagon-triangle topology depicted in \reffi{fig:pentagon-triangle}.
In QCD, the integrals of this topology have at most rank $6$ in total, appearing in the three maximal rank combinations $(5, 1)$, $(4, 2)$, and $(3, 3)$ in $(q_1, q_2)$.
Reducing these three integrals of maximal ranks separately to rank $0$ results in the runtimes shown in \refta{tab:runtimes}, where we did not include the $\tilq^2$ terms as discussed before.
\begin{table}[htbp]
  \caption{
    Runtimes to reduce the three maximal-rank integrals of the pentagon-triangle topology to rank $0$.
  }
  \label{tab:runtimes}
  \begin{center}
    \begin{tabular}{c|c|c}
      \toprule
      Rank & Tensor integral mode & Amplitude mode \\
      \midrule
      $(5, 1)$ & $2.7$\,ms per psp & $0.51$\,ms per psp \\
      $(4, 2)$ & $37$\,ms per psp & $0.35$\,ms per psp \\
      $(3, 3)$ & $53$\,ms per psp & $0.36$\,ms per psp \\
      \bottomrule
    \end{tabular}
  \end{center}
\end{table}
While all runtimes are already promisingly low in the millisecond regime, we notice that the amplitude mode significantly outperforms the tensor integral mode, just as we expected based on the theoretical arguments in \refse{se:red2l}.
Further optimisations of the code leave room for an even better performance in the future.

\begin{figure}[htbp]
  \centering
  \includegraphics[width=0.6\textwidth]{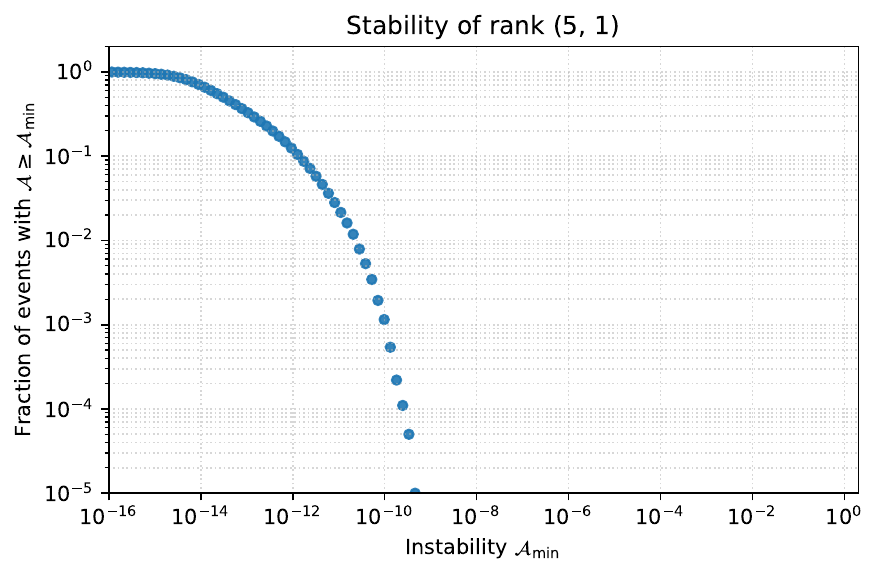}
  \caption{
    Stability of the recursive reduction of the rank-$(5, 1)$ pentagon-triangle integral to ranks $(0, 1)$ and $(1, 1)$.
  }
  \label{fig:instability}
\end{figure}

Since \refeq{eq:TIdia2l-red_rank1} is valid on the \emph{integrand} level, we can verify the recursive reduction to ranks $1$ and $0$ by simply evaluating the \emph{integrands} directly at randomly chosen numerical points for the external kinematics and the loop momenta.
We define the instability of the recursive integrand-level reduction as
\begin{equation}
  \mathcal{A} = \left|\frac{I_{\text{recursive}} - I_{\text{direct}}}{I_{\text{direct}}}\right| ,
\end{equation}
where $I_{\text{recursive}}$ denotes evaluating \refeq{eq:TIdia2l-red_rank1} and $I_{\text{direct}}$ a direct evaluation of the integrand before reduction for fixed values of $q_1$ and $q_2$, both in double precision.
To avoid infrared-singular regions, we apply the cut $p_{a,\text{T}} > 0.05 \sqrt{s}$ on the transverse momenta of the final-state particles.
In \reffi{fig:instability} we show the fraction of points with an instability $\mathcal{A} \geq \mathcal{A}_{\mathrm{min}}$ for a sample of $100\,000$ uniform random phase space points at $\sqrt{s} = 1\,$TeV.
We see that all of the points achieve an accuracy of more than eight digits.
Around $99.9\,$\% of the points even achieve an accuracy of at least nine digits.

The validation and measurement of the numerical stability of the final integral-level reduction steps require also the reduction and evaluation of the scalar integrals.
This will be discussed in a forthcoming publication~\cite{Lange:2026XXX}.

\section{Conclusions}

In these proceedings we presented a new, general algorithm to reduce arbitrary two-loop tensor integrals to scalar integrals numerically.
It allows us to reduce the integrals recursively to rank $1$ in each of the loop momenta, which makes it particularly well-suited for a numeric implementation in the \OpenLoops{} framework.
The remaining integrals can then be reduced to scalar integrals with a standard covariant decomposition, which is reasonably cheap due to the small systems of equations remaining at low ranks.
We verified and tested our preliminary implementation at the hand of a $2\to2$ topology and observed promising runtimes and stability.

To complete our proof-of-concept code, the next goal is to implement an interface to the program \texttt{Kira}~\cite{Maierhofer:2017gsa,Klappert:2020nbg,Lange:2025fba} to reduce the scalar integrals to a basis of master integrals, and afterwards an interface to existing tools to compute the master integrals, see e.g.\ \citeres{Smirnov:2021rhf,Liu:2022chg,Heinrich:2023til}.
In this step we will then also include the $\tilq^2$ integrals.
More details will be discussed in a forthcoming publication~\cite{Lange:2026XXX}.

\acknowledgments

This work was supported by the Swiss National Science Foundation (SNSF) under contract \href{https://data.snf.ch/grants/grant/211209}{TMSGI2\_211209}.

\bibliographystyle{JHEP}
\bibliography{bib}

\end{document}